\documentclass[aps,prb,twocolumn,amsmath,amssymb,superscriptaddress,floatfix]{revtex4}
\usepackage{graphicx}
\usepackage{bm}
\usepackage{amssymb}
\usepackage{amsmath}
\usepackage{color}
\usepackage{ulem}
\newcommand{\be}{\begin{equation}}
\newcommand{\ee}{\end{equation}}
\newcommand{\beqn}{\begin{eqnarray}}
\newcommand{\eeqn}{\end{eqnarray}}

\begin{document}

\title{Phase-transitions of the random bond Potts chain with long-range interactions}

\author{Jean-Christian Angl\`es d'Auriac}
\affiliation{Institut N\'eel-MCBT CNRS, B. P. 166, F-38042 Grenoble, France}
\author{Ferenc Igl\'oi}
\email{igloi.ferenc@wigner.mta.hu}
\affiliation{Wigner Research Centre, Institute for Solid State Physics and Optics, H-1525 Budapest, P.O.Box 49, Hungary}
\affiliation{Institute of Theoretical Physics, Szeged University, H-6720 Szeged, Hungary}

\date{\today}

\begin{abstract}
We study phase-transitions of the ferromagnetic $q$-state Potts chain with random nearest-neighbour couplings having a variance $\Delta^2$ and with homogeneous long-range interactions, which decay with the distance as a power $r^{-(1+\sigma)}$, $\sigma>0$. In the large-$q$ limit the free-energy of random samples of length $L \le 2048$ is calculated exactly by a combinatorial optimization algorithm. The phase-transition stays first-order for $\sigma < \sigma_c(\Delta) \le 0.5$, while the correlation length becomes divergent at the transition point for $\sigma_c(\Delta) < \sigma < 1$. In the latter regime the average magnetization is continuous for small enough $\Delta$, but for larger $\Delta$ it is discontinuous at the transition point, thus the phase-transition is of mixed order.
\end{abstract}

\maketitle
\section{Introduction}
\label{sec:intr}
The properties of phase-transitions in a pure system can be modified due to quenched disorder. This problem has been studied in most
details at a second-order transition point\cite{harris,ccfs}, but much less is known when the transition is of first order\cite{Cardy99}. 
When the disorder is coupled to the local energy density, such as for bond disorder, there is a general tendency that
the latent heat at the transition point is reduced\cite{imrywortis}. In two-dimensional systems with nearest-neighbour (or short-range (SR)) interactions
any amount of bond disorder is enough to turn the transition into second order\cite{aizenmanwehr}. The new universality class of the problem, however, remains
unknown and numerical investigations are needed to identify the properties of the emergent random fixed point\cite{pottsmc,pottstm,jacobsenpicco,ai03,long2d}. In three- and higher dimensional
SR systems, however, weak disorder is generally irrelevant, thus the phase-transition stays discontinuous and only for strong enough disorder
will it turn to a second-order one. This type of problem has been numerically studied for the $q$-state Potts model with $q>2$\cite{uzelac,pottssite,pottsbond,mai05,mai06}. In particular
a mapping between the random-field Ising model (RFIM) and the Potts model in the $q \to \infty$ limit has been used to predict some tricritical
exponents of the latter random model\cite{pottstm,mai06}.

Homogeneous, i.e. nonrandom systems with long-range (LR) interactions could have an ordered phase\cite{dyson} and a
first-order transition, too, even if the system is one-dimensional. This happens, among others for the $q$-state Potts chain\cite{Wu} with
power-law interactions
\be
J(r) \approx J r^{-(1+\sigma)}\;,
\label{J(r)}
\ee
where $r$ is the distance between the sites and the exponent is
$\sigma>0$ to have extensive total energy (for $\sigma<0$ one should divide $J$ by $L^\sigma$).
According to numerical results\cite{pure} the transition in the LR Potts chain is of
first order for sufficiently large values of $q$, where the limiting value $q_c=q_c(\sigma)$ is an increasing function of $\sigma$.
On the other hand the transition for $q<q_c(\sigma)$ is of second order.

Low-dimensional LR models with power-law interactions became the subject of intensive research recently, after
it has been noticed that the decay exponent, $\sigma$ in the problems plays the role of some kind of effective dimensionality
of the analogous SR model. Among the classical problems studied so far we mention the non-random Ising model in one- and two-dimensions\cite{fisher_me,sak,bloete,picco,parisi}, the
non-random Potts chain\cite{pure}, the Ising spin-glass model\cite{kotliar,cecile_sg,cecile_sg1,cecile_sg2} and the RFIM in one dimension\cite{bray,weir,rodgers,aizenman,cassandro,monthus11,leuzzi,dewenter}. For quantum models we mention investigations of
the transverse-field Ising model both with pure\cite{porras,deng,hauke,peter,nebendahl,wall,cannas,dutta,dalmonte,koffel,hauke1}
and random couplings\cite{jki,cecile,jki1,kji} and the Anderson localization problem\cite{cecile_anderson},
for reaction-diffusion type models the contact process and similar models with\cite{jki1} and without\cite{mollison,janssen,howard,ginelli,fiore,linder,grassberger,ginelli1,adamek,hinrichsen}
quenched disorder.

The critical properties of LR models are often unusual. Here we mention that the classical Ising chain for $\sigma=1$, as well as other one-dimensional discrete spin models with LR interaction have a so-called
mixed-order (MO) phase transition\cite{anderson1969exact,thouless1969long,dyson1971ising,cardy1981one,aizenman1988discontinuity,slurink1983roughening,bar2014mixed}, at which point the order-parameter has a jump, but at the same time the correlation length is divergent.
We note that recently MO transitions have been observed in other problems, too\cite{PS1966,fisher1966effect,blossey1995diverging,fisher1984walks,gross1985mean,toninelli2006jamming,toninelli2007toninelli,schwarz2006onset,liu2012core,liu2012extraordinary,zia2012extraordinary,tian2012nature,bizhani2012discontinuous,sheinman2014discontinuous,bar2014mixed1,kji}.

In the present paper we consider LR models, having a first-order transition in their non-random version and study the effect of
quenched disorder on the phase-transition properties of the system. To be specific, we consider the LR Potts model in
one dimension for large values of $q$ (actually we consider the $q \to \infty$ limit), when the transition of the pure model
is of first order for all values of the decay exponents, $\sigma > 0$. We have random nearest-neighbour couplings with a variance
$\Delta^2$, but the long-range forces are non-random and follow the behaviour in Eq.(\ref{J(r)}).
We study the phase-transition of the system for different values of the effective dimensionality ($\sigma$) and the
strength of disorder ($\Delta$). The free energy and the magnetization of a given random sample is calculated exactly
by a computer algorithm, which works in polynomial time\cite{aips02}. We follow the temperature dependence of the average magnetization in relatively large
finite samples and the location of the phase-transition point and its properties are analyzed by finite-size extrapolation methods.

The structure of the paper is the following. The model and some results are summarized in Sec.\ref{sec:model}. Numerical results at different
points of the phase diagram are presented in Sec.\ref{sec:numerical} and analyzed by finite-size scaling. We close our paper with a
discussion in Sec.\ref{sec:discussion}.

\section{Model and some results}
\label{sec:model}

We consider the ferromagnetic $q$-state Potts-model\cite{Wu} in a one-dimensional periodic lattice with long-range interactions defined by the Hamiltonian:
\be
{\cal H}=-\sum_i J_i \delta(s_i,s_{i+1})-\sum_{i < j+1} J_{ij} \delta(s_i,s_{j})\;.
\label{hamiltonian}
\ee
Here $s_i=1,2,\dots,q$ is a Potts-spin variable at site $i=1,2,\dots,L$ and the long-range interaction, $J_{ij}$, has a
power-law dependence as in Eq.(\ref{J(r)}) with
$r=\min\left(L-\left|i-j\right|,\left|i-j\right|\right)$.
The nearest neighbour couplings, $J_i \equiv J_{i,i+1}$, are random variables.
For simplicity we take $J_i$ from a bimodal distribution, being either $J_-=J-\Delta$ or $J_+=J+\Delta$ with
equal probability. In the following we set the energy-scale to $J=1$ and restrict ourselves to $0<\Delta<1$.

\subsection{The large-$q$ limit}
\label{sec:large-q}

In this paper we consider the $q \to \infty$ limit of the model, when the reduced free-energy in the Fortuin-Kasteleyn representation\cite{kasteleyn}
is dominated by a single graph\cite{JRI01}, the so called optimal graph, $G$, and given by:
\be
-\beta f L= {\rm max}_G W(G),\quad W(G)= \left[c(G) + \beta \sum_{ij \in G} J_{ij} \right]\;,
\label{max}
\ee
Here $c(G)$ stands for the number of connected components of $G$ and $\beta=1/(T \ln q)$, with the temperature $T$.

In the \textit{homogeneous nonrandom model} with $\Delta=0$ there are only trivial optimal
graphs as shown in appendix for any $\sigma$. In the low-temperature phase, $T < T_c$, it is the
fully connected graph with $W(G_c)=-\beta L f_{\rm hom}=1+L \beta {\cal Z}_L(1+\sigma)$, with
\beqn
{\cal Z}_L(1+\sigma)&=&\frac{1}{L}\sum_{i=1}^{L/2}(L+1-i)i^{-(1+\sigma)} \nonumber \\
&=&\left(1+\frac{1}{L}\right) \zeta_{L/2}(1+\sigma)-\frac{1}{L}\zeta_{L/2}(\sigma)\;,
\label{zeta}
\eeqn
where we have assumed that $L$ is \textit{even}.
Here $\zeta_{L/2}(\alpha)=\sum_{i=1}^{L/2} i^{-\alpha}$ and for $L \to \infty$ we have the Riemann-zeta function, $\zeta(\alpha)$.
In the high-temperature phase, $T > T_c$, the optimal graph is the empty graph with $W(G_e)=-\beta L f_{\rm hom}=L$. The phase-transition point
in the
thermodynamic limit is given by $\beta_c=1/\zeta(1+\sigma)$ where the phase transition is of first order having the maximal jump in
the magnetization.

In the limit where $\sigma$ goes to infinity one recovers the disordered SR Potts chain.
In that case and for finite size $L$, there are non trivial optimal sets. But in the
thermodynamical limit, the magnetization still jumps from zero to one for the bimodal
distribution. This is shown in the appendix.

\subsection{Stability analysis of the random model}
\label{sec:stability}

Here we start with weak disorder, $\Delta \ll 1$, and estimate the characteristic function of non-homogeneous optimal graphs. First let us
consider an island of $l +1 \le \frac{L}{2}$ consecutive sites, which are fully connected within the sea of isolated points. The corresponding characteristic
function is given by:
\be
W(G_1)=L-l+\beta [(l+1) \zeta_l(1+\sigma)-\zeta_l(\sigma)] + \beta \Delta \epsilon(l)\;,
\ee
where $\epsilon(n)$ is the sum of $n$ random numbers with mean zero and variance unity, thus $\epsilon(n) \sim \sqrt{n}$ for large $n$.
At the transition point of the pure system, $\beta=\beta_c=1/\zeta(1+\sigma)$, the new diagram is the optimal set, i.e. $W(G_1)>W(G_e)$,
provided: $\Delta > [l (\zeta(1+\sigma)-\zeta_{l-1}(1+\sigma))+\zeta_l(\sigma)]/\epsilon(l-1)$. For large-$l$ the r.h.s. of this
inequality scales as: $l^{1-\sigma}/l^{1/2} \sim l^{1/2-\sigma}$, thus we have the condition 
\be
\Delta > C l^{1/2-\sigma},\quad l \gg 1\;.
\label{Delta}
\ee
Consequently for a decay exponent $\sigma > 1/2$ there is a new, non-homogeneous optimal set and the (phase-transition) properties of the system
are modified by any small amount of disorder, at least in the thermodynamic limit.
On the contrary for $\sigma < 1/2$ the transition, at least for small $\Delta$ stays
first order and it could be changed only by strong enough disorder, i.e. for large $\Delta$.

Next we study the stability of the fully connected graph $G_c$ and consider a diagram, $G_2$, in which in a fully connected sea of points there are $l$ disconnected sites. Its
characteristic function is given by: 
%
\beqn
&W(G_2)=l+1+\beta L {\cal Z}_L(1+\sigma) \\
&-\beta[ l (2\zeta_{L/2}(1+\sigma)-\zeta_{l}(1+\sigma))+\zeta_l(\sigma)]
+ \beta \Delta \epsilon(l+1) \nonumber \;.
\eeqn
At $\beta=\beta_c$ we have $W(G_2)<W(G_c)$, at least for weak disorder for any value of $\sigma > 0$. This means, that considering the stability of the two trivial optimal sets of the pure system at $\beta=\beta_c$, these are not symmetric. For $\sigma>1/2$ the empty diagram is unstable, while the fully connected graph is stable for weak disorder. We note that in the SR model both graphs become unstable at the same value of the dimensionality: $d \le 2$.

In the LR model in the modified transition regime, $\sigma>1/2$, we can define a breaking-up length:
\be
l^{*} \sim \Delta^{1/(1/2-\sigma)},\quad \sigma>1/2\;,
\ee
which is the typical size of connected clusters. This means, that in a finite system one should have $L > l^{*}$ in order to
be able to observe a new type of transition, otherwise there is a pseudo-first-order transitions in the finite system.

\subsection{Relation with the RFIM}
\label{sec:random}

The previous stability analysis is based on the properties of an interface separating the two trivial
optimal graphs and analogous reasoning due to Imry and Ma\cite{imry_ma} works for the RFIM, in which case the interface separates the ordered
and disordered regions of the model. This mapping has been observed by Cardy and Jacobsen\cite{pottstm} and can be generalized
for LR interactions in which case the RFIM in a one-dimensional lattice is
defined by the Hamiltonian:
\be
{\cal H}_{RFIM}=-\sum_i B_i S_i-\sum_{i < j} J_{ij} S_iS_{j}\;,
\label{RFIM}
\ee
in terms of $S_i=\pm 1$. Here $B_i$ is a random variable with zero mean and variance $\Delta^2$ and $J_{ij}$ is in the same form
as in Eq.(\ref{J(r)}). The critical behavior of ${\cal H}_{RFIM}$ has been studied in the
literature\cite{bray,weir,rodgers,aizenman,cassandro,monthus11,leuzzi,dewenter} and $\sigma$-dependent properties are found, which are summarized in the following.

There is a ferromagnetic ordered phase in the system for $0<\sigma<1/2$ (which corresponds to phase-coexistence, i.e.
first-order transition in the RBPM) and there is no spontaneous ordering for $\sigma > 1/2$ (which is analogous to the absence of
first-order transition in the RBPM). The transition to the ferromagnetic ordered phase is
mean-field (MF) type in the region $0<\sigma<1/3$, where
the critical exponents are the MF ones: $\alpha_{RF}=0$, $\beta_{RF}=1/2$, $\gamma_{RF}=1$ and $\nu_{RF}=1/\sigma$.
On the contrary for $1/3<\sigma<1/2$ the transition is non-MF:
the critical exponent $\nu_{RF}$ is not known exactly, but we have the relations:
\be
\frac{2-\alpha_{RF}}{\nu_{RF}}=1-\sigma,\quad\frac{\beta_{RF}}{\nu_{RF}}=\frac{1}{2}-\sigma,\quad \frac{\gamma_{RF}}{\nu_{RF}}=\sigma
\ee
Cardy and Jacobsen\cite{pottstm} has conjectured relations between the magnetization exponents of the RFIM
and the tricritical exponents in the energy sector of the RBPM, at least for SR models. If we assume the validity
of these relations for LR interactions, too, we have for the correlation-length exponent of the RBPM at the tricritical point:
\be
\nu=\frac{\nu_{RF}}{\beta_{RF}+\gamma_{RF}}\;.
\ee
Thus the conjectured results are $\nu=\dfrac{2}{3\sigma}$ and $\nu=2$ in the MF-region and in the non-MF region, respectively.

\section{Numerical calculation}
\label{sec:numerical}

\subsection{Preliminaries}
\label{sec:preliminaries}

As for systems with quenched disorder one should perform two averages: first, the thermal average for a given realization of
disorder and second, averaging over the disorder realizations. For a given random sample of length $L$ the thermal average is obtained through
the solution of the optimization problem given in Eq.(\ref{max}). Having the optimal graph of the sample, we have the free-energy
as well as the structure of connected clusters in this graph. The magnetization of the sample, $m$, is given by the number of sites
in the largest cluster, $N_{\rm max}$, as $m=N_{\rm max}/L$. The optimization process for a given sample is solved exactly by
a combinatorial optimization algorithm which works polynomially in time\cite{aips02}. This makes us possible to treat relatively large
samples up to $L=1024$ and in some cases up to $L=2048$. In the latter case the typical computational time of a sample in the complete
temperature range is
about 6-7 hours in a 2.4 GHz processor. A drawback of the calculation, that the possible graphs in the present
problem are fully connected, having $L(L-1)/2$ possible edges and the algorithm needs so many iterations, which increases the computational
time accordingly. In the second step of the averaging process we have considered several independent random samples, their
typical number being a few $10000$, for $L=1024$ a few $1000$.

\subsection{Magnetization profiles}
\label{sec:profiles}

\begin{figure}
  \begin{center}
    \includegraphics[width=9cm]{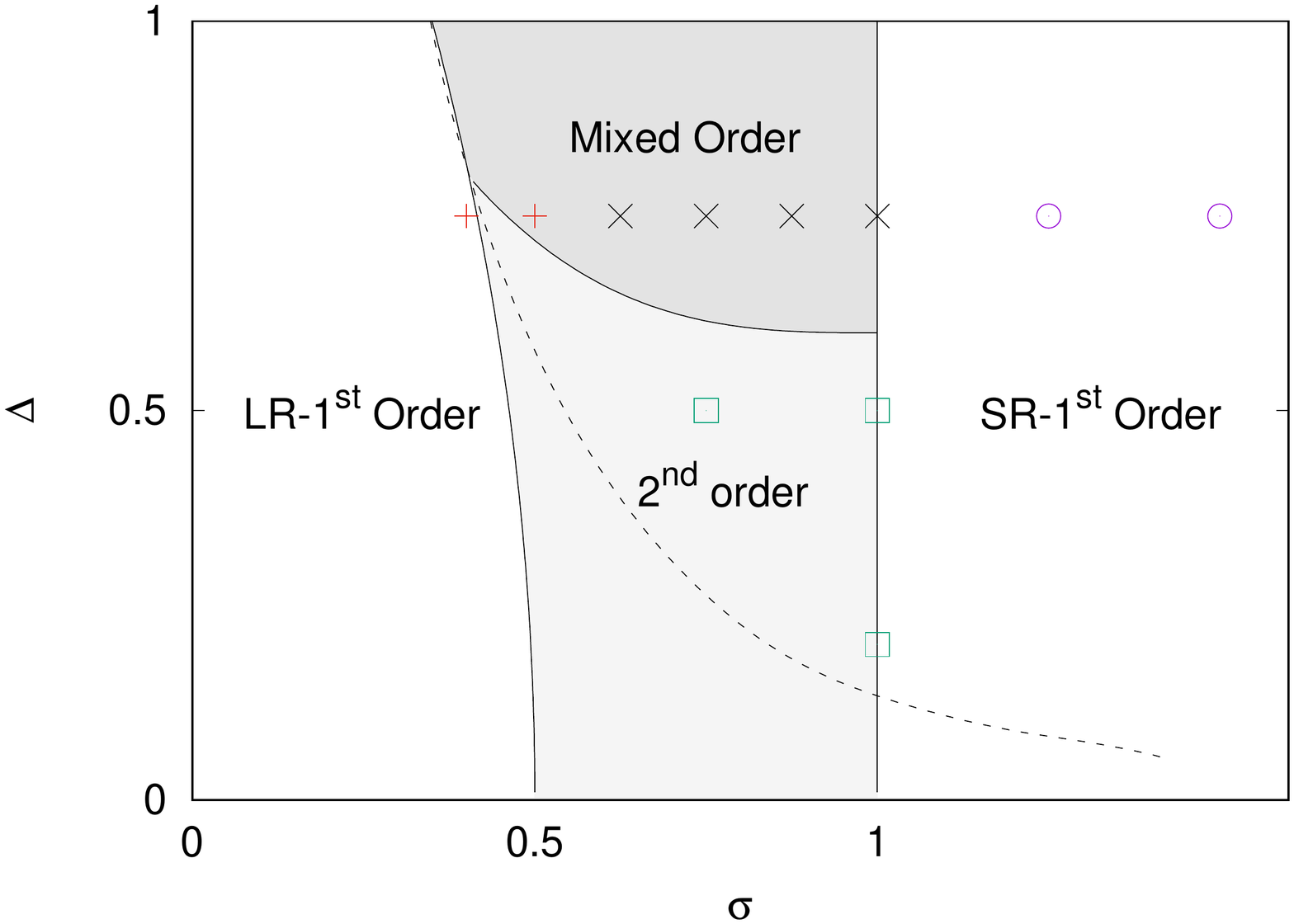}
    
  \end{center}
  \caption{Schematic phase-diagram of the LR Potts chain with random nearest neighbour couplings in the $q \to \infty$ limit together with points of the phase diagram studied numerically. The border of the strongly $1^{\rm st}$ order regime in a finite system (dashed line) is calculated with $L=256$ and with $N^{\#}=600$ samples, see text. The plus sign refers to long range first order, the cross sign refers to mixed order, the circle to short range first order, and the square to second order transitions, respectively. }
  \label{fig:phasediag}
\end{figure}
Before entering in details to study the phase-diagram of the system we have made a rough estimate of the domain, in which the transition is
very strongly first order. For this purpose we have analysed the phase-transition of $N^{\#}=600$ samples of length $L=256$. In the
shaded area of the $\sigma - \Delta$ phase diagram in Fig.\ref{fig:phasediag} in all samples the transition is between the fully connected graph and the empty graph, thus the transition is maximally $1^{\rm st}$ order, as in the homogeneous system. Then,
we have chosen a several points outside the strongly first-order regime, which are indicated in Fig.\ref{fig:phasediag}.
The selected points can be devided into two groups: a set of points with relatively weak disorder, $\Delta=0.2$ and $\Delta=0.5$ and
another set with quite strong disorder $\Delta=0.75$.
At each point the calculation of the optimal graph is performed in the complete temperature range: we have monitored the temperature
dependence of the magnetization and focused to its possible singular behaviour. These calculations are performed in finite systems
with L=64,128,\dots 1024 and the actual properties of the singularity, thus the
form of the phase transition is analyzed by finite-size extrapolation.

\subsubsection{Weak and intermediate disorder regimes: $\Delta=0.2$ and $0.5$}
\label{sec:weaker}

\begin{figure}
  \begin{center}
    \includegraphics[width=9cm]{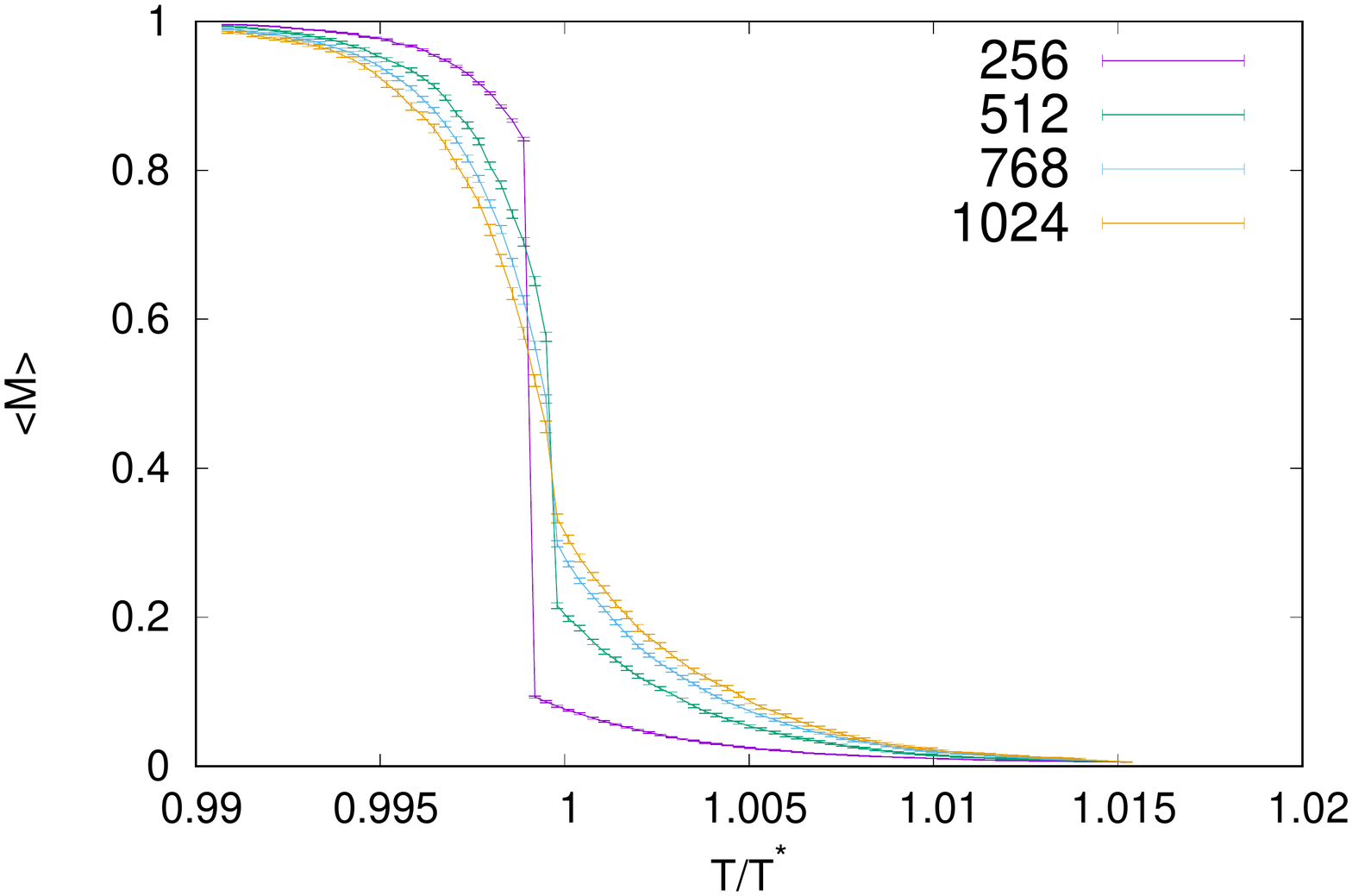}
  \end{center}
  \caption{Second-order transition at $\Delta=0.2$ and $\sigma=1$}
  \label{fig:M_0.2_1}
\end{figure}

\begin{figure}
  \begin{center}
    \includegraphics[width=9cm]{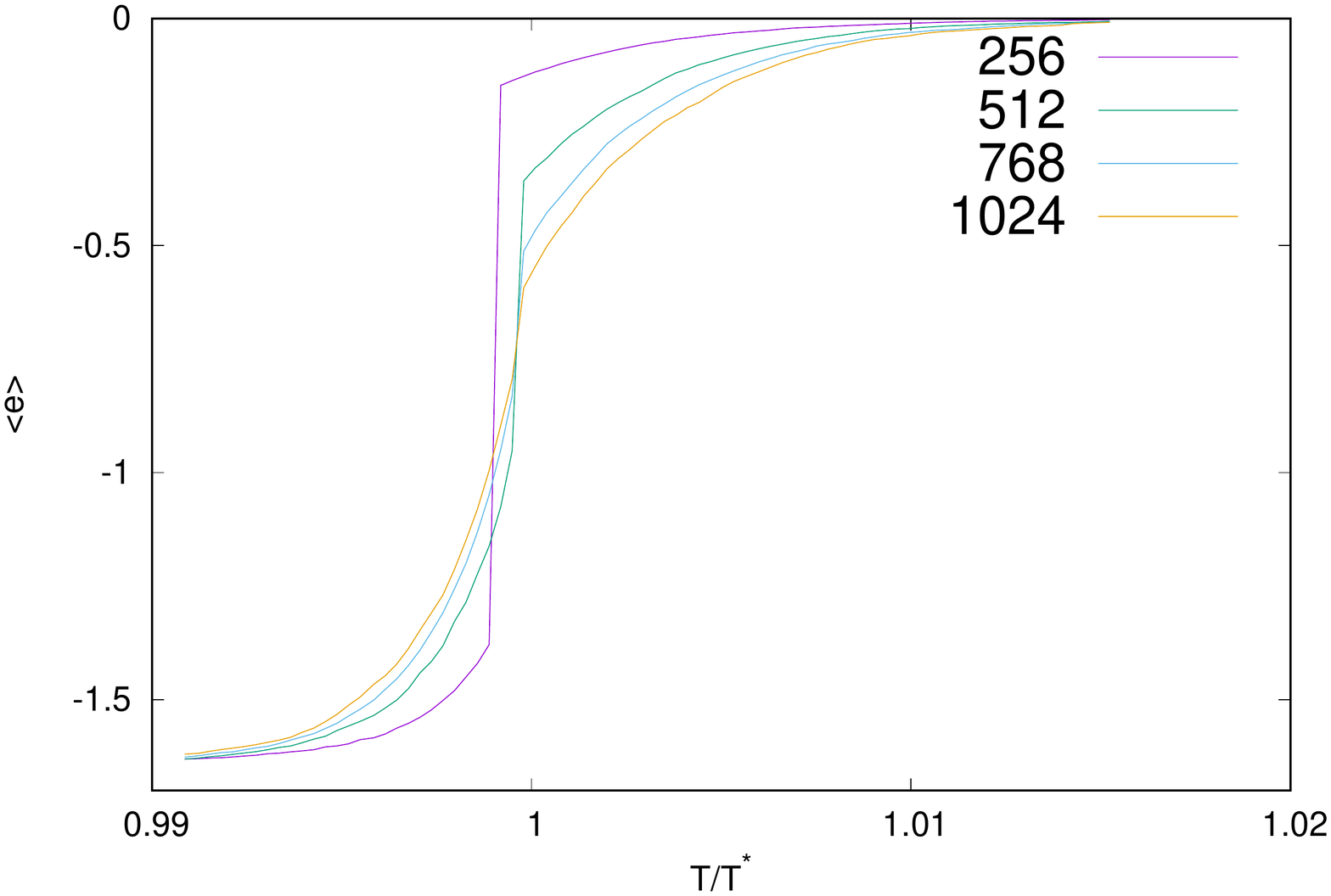}
  \end{center}
  \caption{Energy for $\Delta=0.2$ and $\sigma=1$}
  \label{fig:E_0.2_1}
\end{figure}

For weak disorder with $\Delta=0.2$ we have studied the point of the phase-diagram with $\sigma=1$ (square on Fig.\ref{fig:phasediag}), i.e. at border of the LR regime,
the magnetization profiles are shown in Fig.\ref{fig:M_0.2_1}. It is seen that due to disorder the first-order transition in the pure system is
rounded: the jump in the magnetization is decreasing with increasing size and in the thermodynamic limit the jump is expected to
disappear, $\Delta(L) \sim L^{-\beta/\nu}$, so that the limiting curve $\lim_{L \to \infty} m(L,T)=m(T)$ is continuous. However, its derivative
at $T=T_c$ is expected to be divergent, so that $m(T)-m(T_c) \sim |T-T_c|^{\beta}$. The finite-size transition points are shifted
as $T_c-T_c(L) \sim L^{-1/\nu}$. Note, that for $T<T_c$ ($T>T_c$) the profiles satisfy $m(L_1,T)>m(L_2,T)$ ($m(L_1,T)<m(L_2,T)$) for $L_1<L_2$
. We have studied also the temperature dependence of the average energy-density,
which is shown in Fig.\ref{fig:E_0.2_1}. At the transition point in small finite systems there is a discontinuity of the energy-density, which seems to dissappear
in the thermodynamic limit, but its first derivative, the specific heat is divergent: $C(T) \sim |T-T_c|^{-\alpha}$. Consequently the transition
according to the Ehrenfest classification is of second-order.

\begin{figure}
  \begin{center}
    \includegraphics[width=9cm]{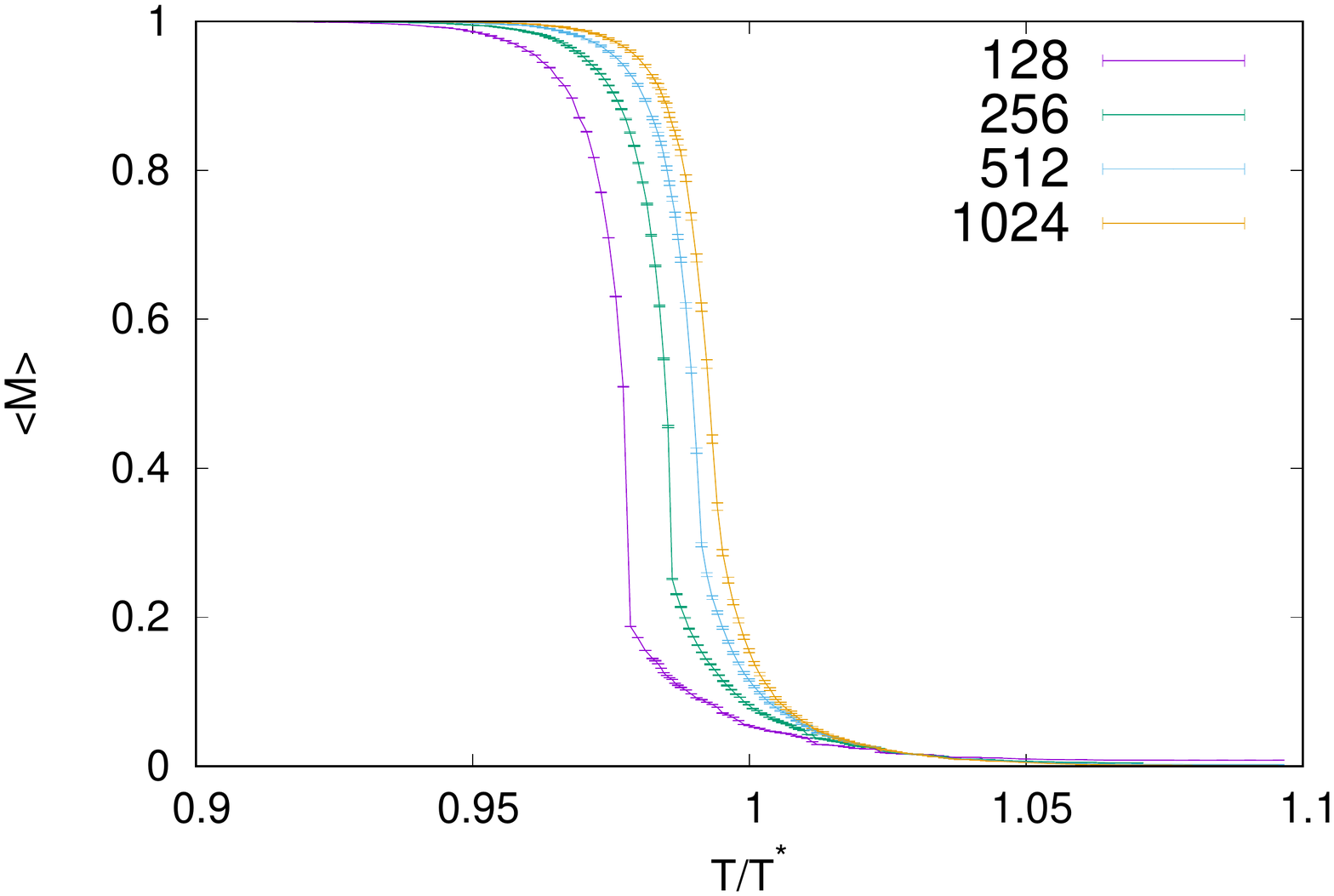}
  \end{center}
  \caption{Second-order transition at $\Delta=0.5$ and $\sigma=0.75$}
  \label{fig:M_0.5_0.75}
\end{figure}

\begin{figure}
  \begin{center}
    \includegraphics[width=9cm]{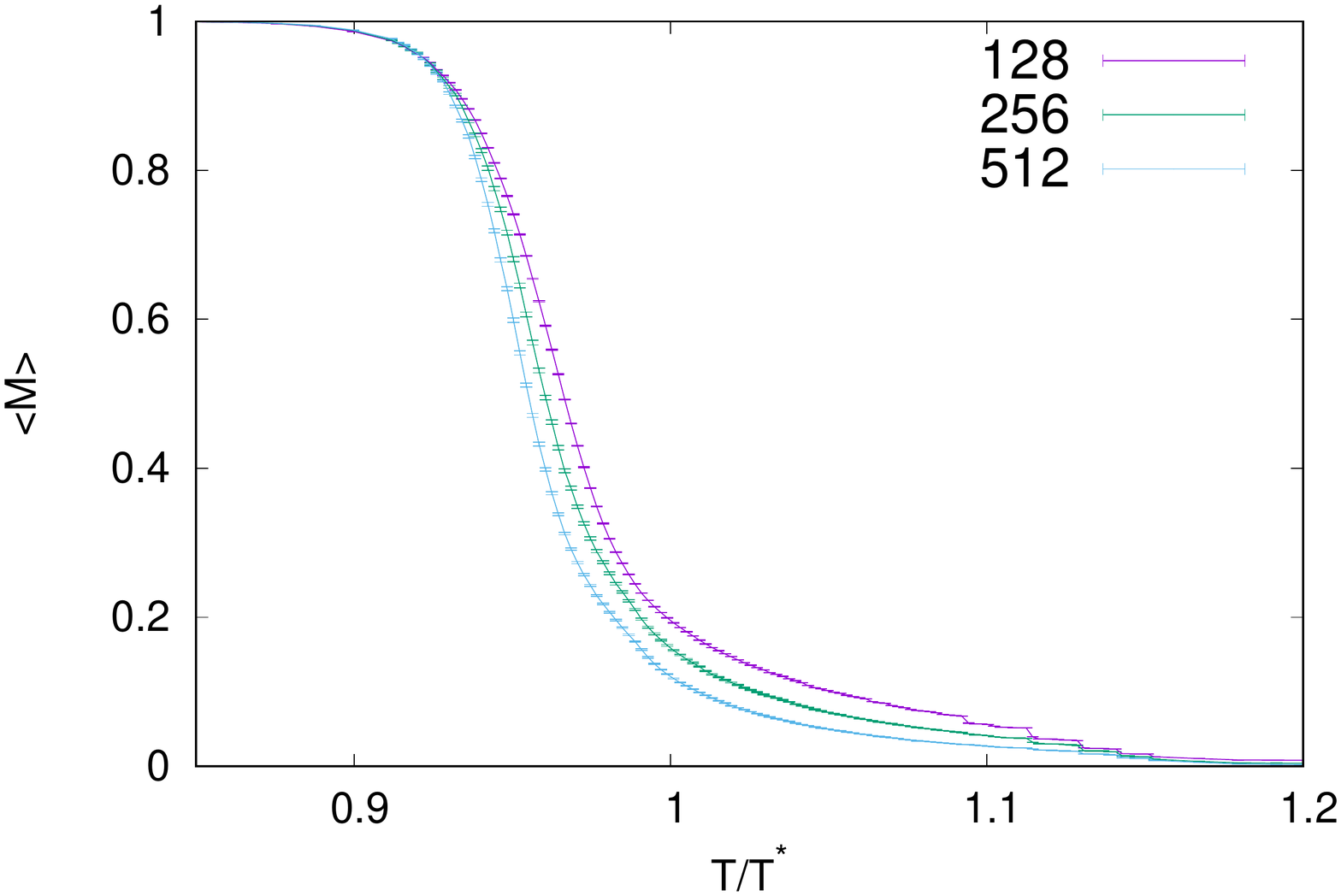}
  \end{center}
  \caption{Magnetisation for $\Delta=0.5$ and at the border $\sigma=1$}
  \label{fig:M_0.5_1}
\end{figure}

For intermediate disorder, $\Delta=0.5$, two points are considered with $\sigma=0.75$ and $\sigma=1.0$, the calculated average magnetization
profiles are presented in Figs. \ref{fig:M_0.5_0.75} and \ref{fig:M_0.5_1}. In both cases the transition seems to be of second-order, which is in agreement with the temperature
dependence of the energy-densities.

\subsubsection{Strong disorder regime: $\Delta=0.75$}
\label{sec:strong}

At the disorder parameter $\Delta=0.75$ we have studied different regimes by varying the decay exponent $\sigma$.

\paragraph{\underline{$\sigma \lesssim 0.5$: LR first-order transitions}}
\label{sec:LR first}

\begin{figure}
  \begin{center}
    \includegraphics[width=9cm]{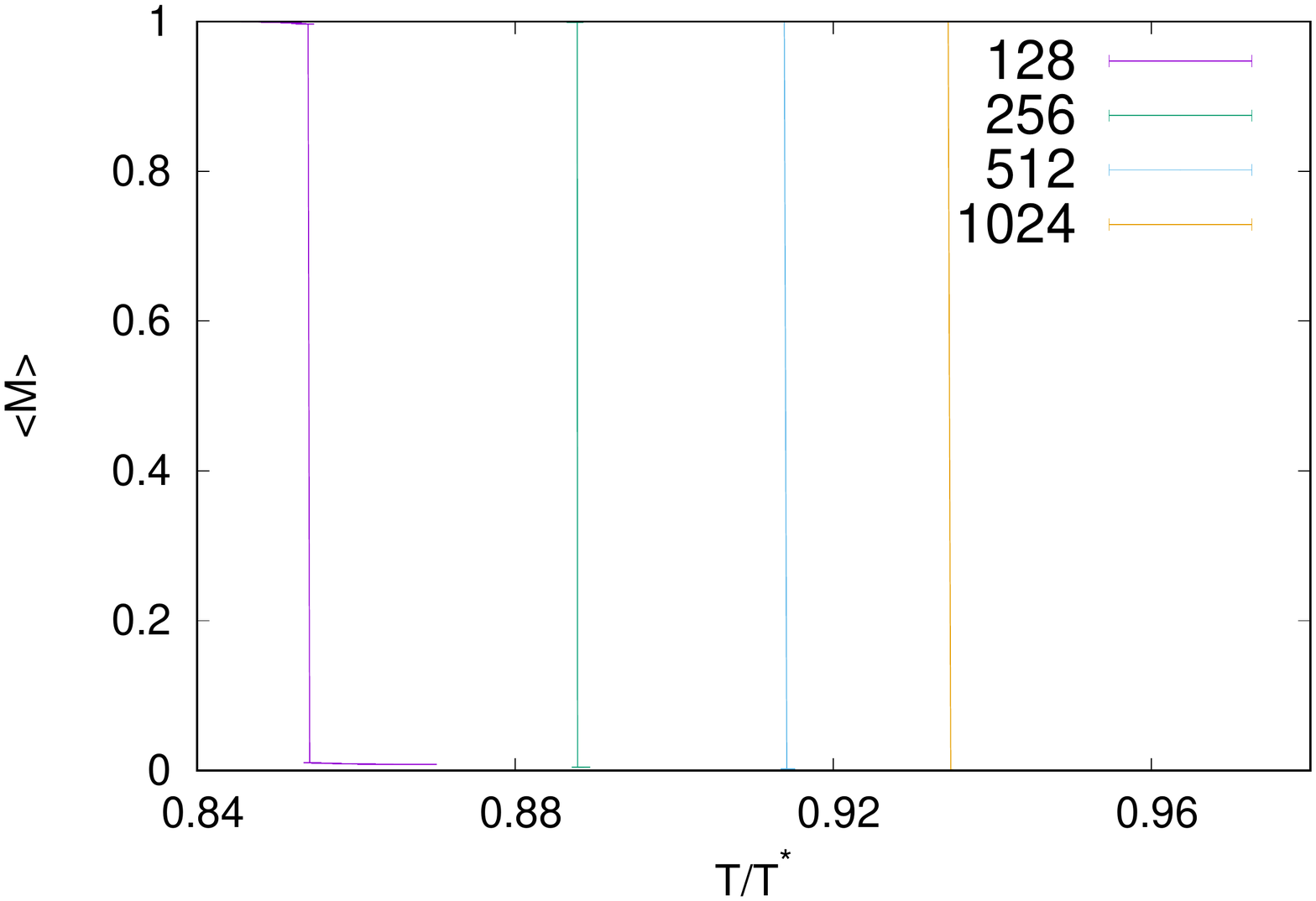}
  \end{center}
  \caption{First-order transition due to LR forces at $\Delta=0.75$ and $\sigma=0.4$.}
  \label{fig:M_0.75_0.4}
\end{figure}

\begin{figure}
  \begin{center}
    \includegraphics[width=9cm]{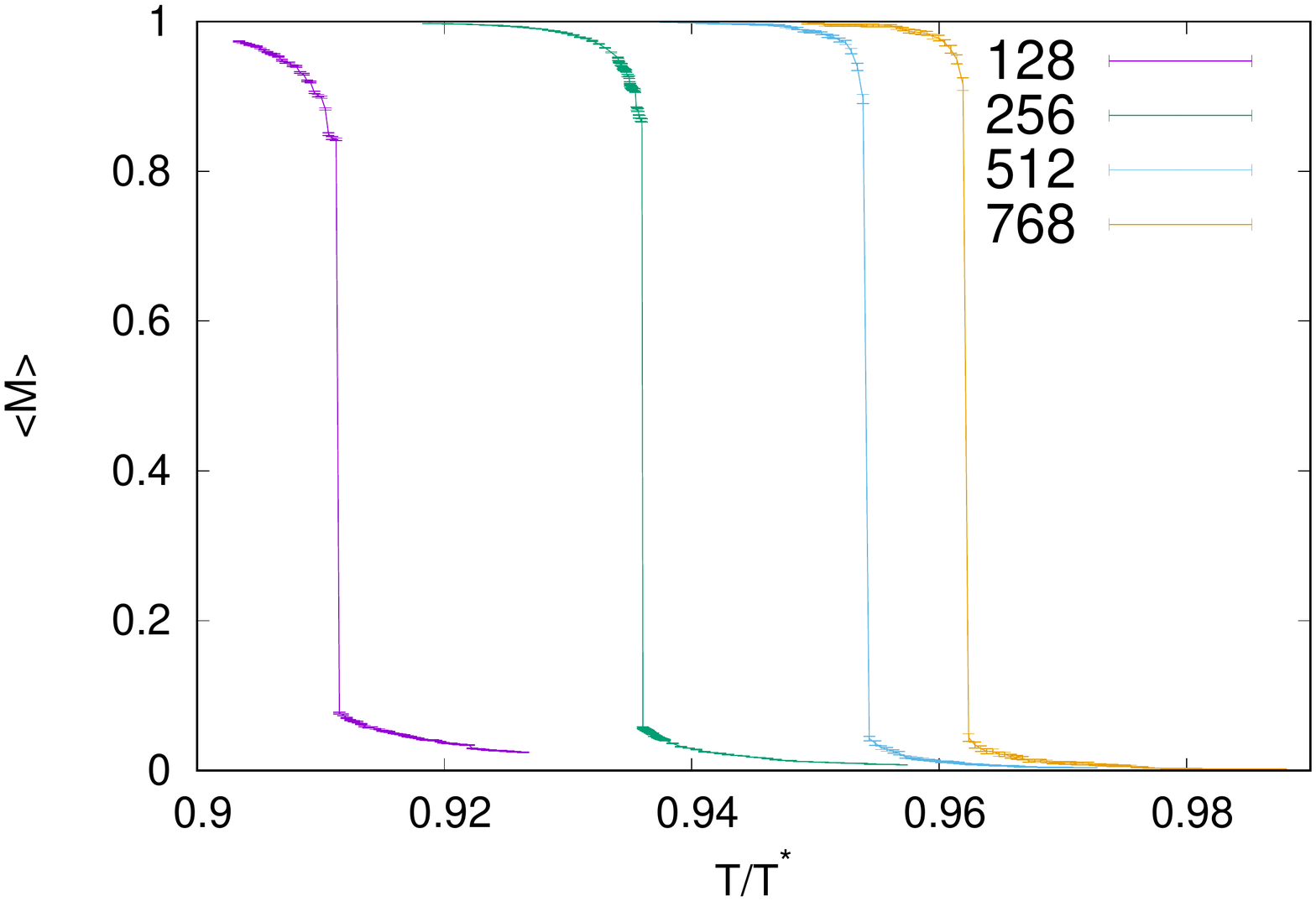}
  \end{center}
  \caption{The jump in the magnetisation is rounded due to disorder at $\Delta=0.75$ and $\sigma=0.5$.}
  \label{fig:M_0.75_0.5}
\end{figure}

\begin{figure}
  \begin{center}
    \includegraphics[width=9cm]{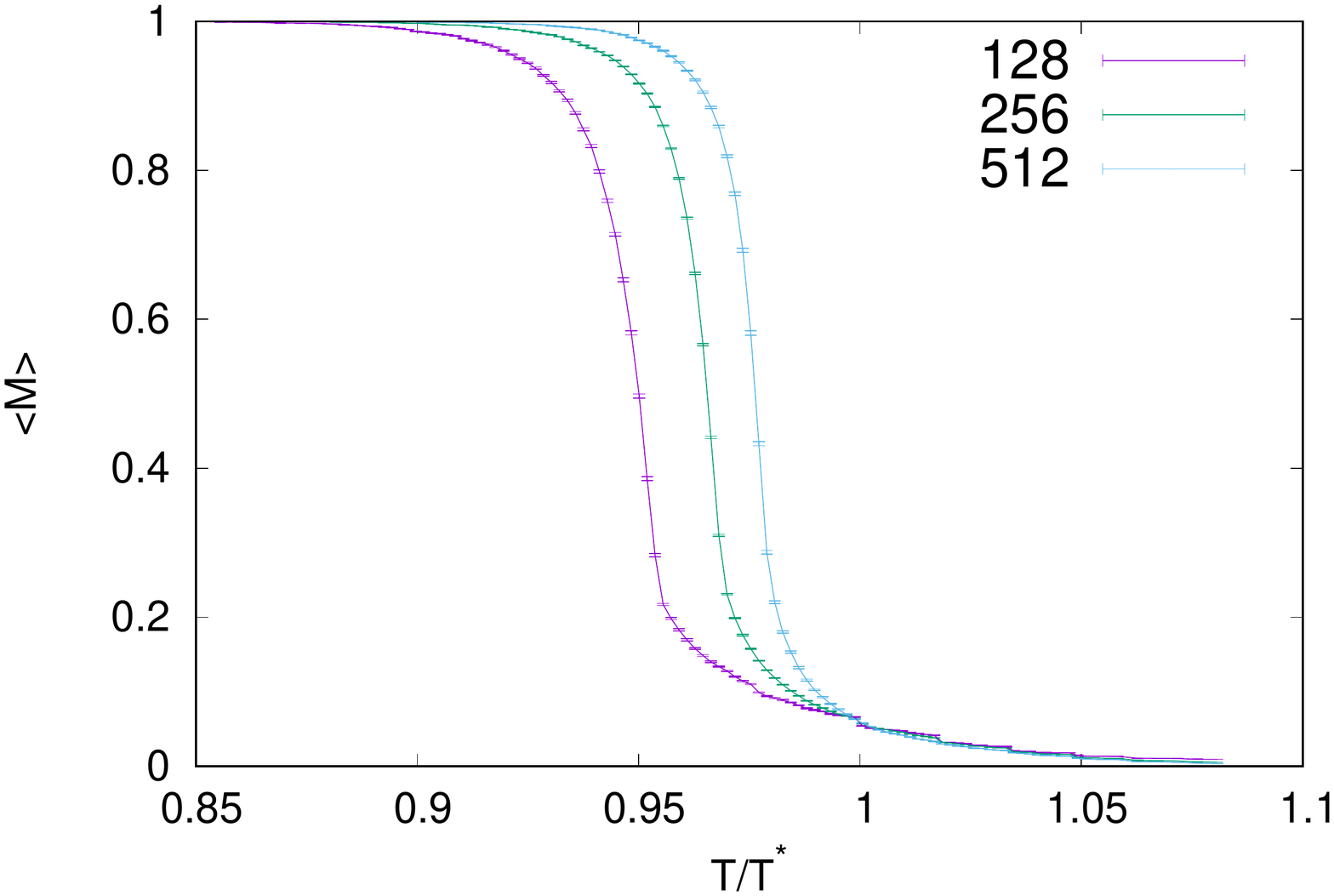}
  \end{center}
  \caption{Mixed-order transition at $\Delta=0.75$ and $\sigma=0.625$}
  \label{fig:M_0.75_0.625}
\end{figure}

\begin{figure}
  \begin{center}
    \includegraphics[width=9cm]{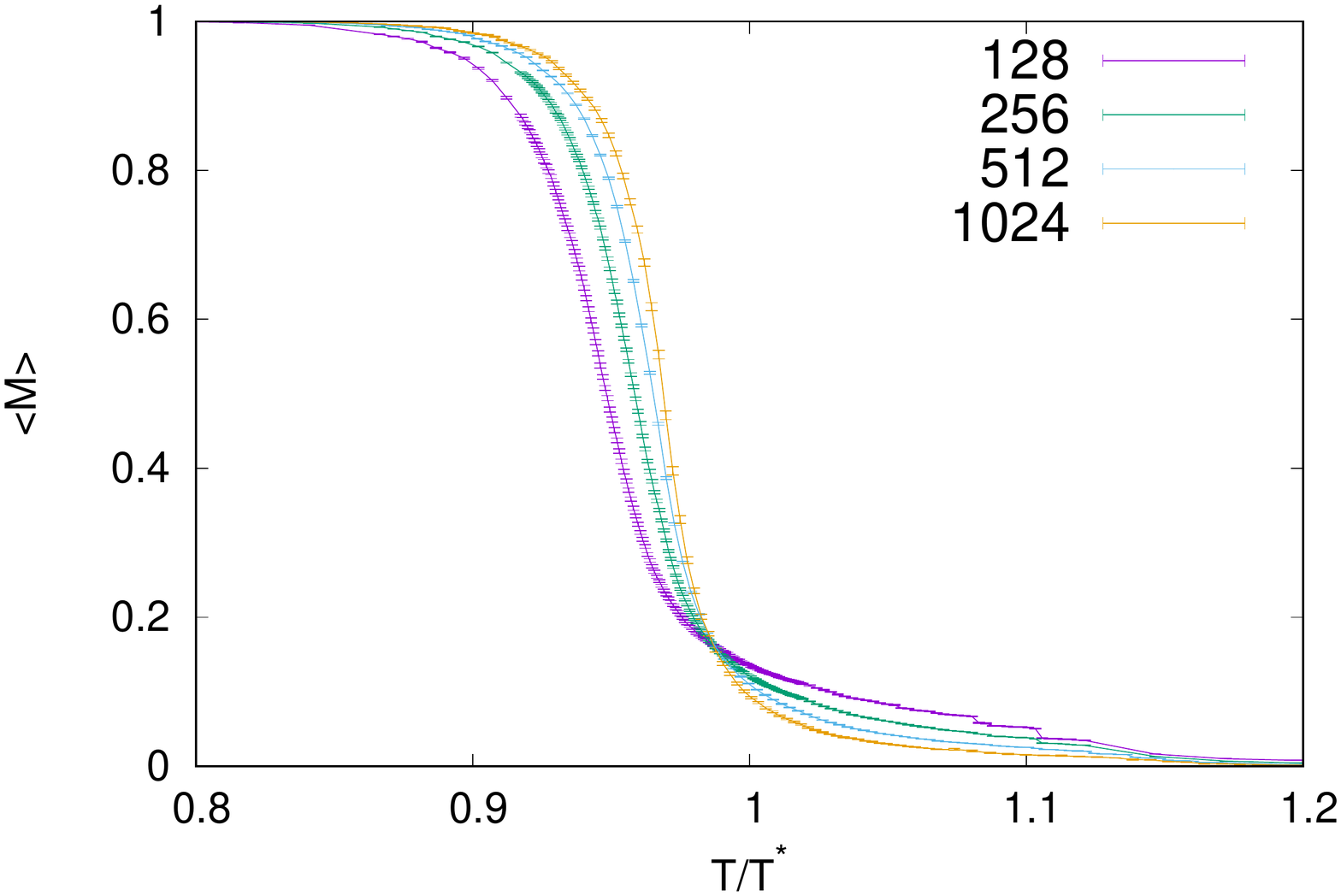}
  \end{center}
  \caption{Mixed-order transition at $\Delta=0.75$ and $\sigma=0.75$}
  \label{fig:M_0.75_0.75}
\end{figure}

\begin{figure}
  \begin{center}
    \includegraphics[width=9cm]{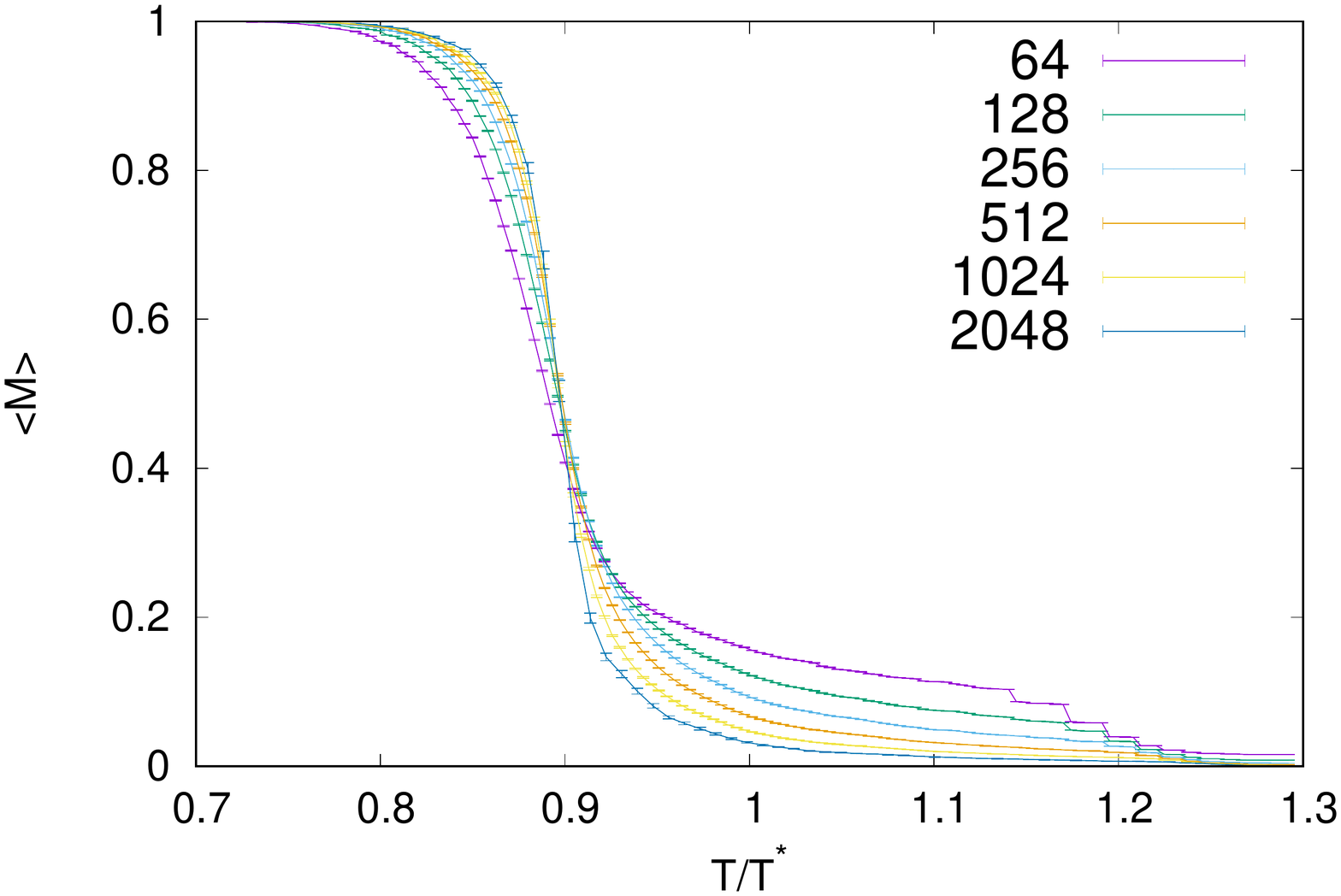}
  \end{center}
  \caption{Mixed-order transition at $\Delta=0.75$ and $\sigma=0.875$}
  \label{fig:M_0.75_0.875}
\end{figure}

\begin{figure}
  \begin{center}
    \includegraphics[width=9cm]{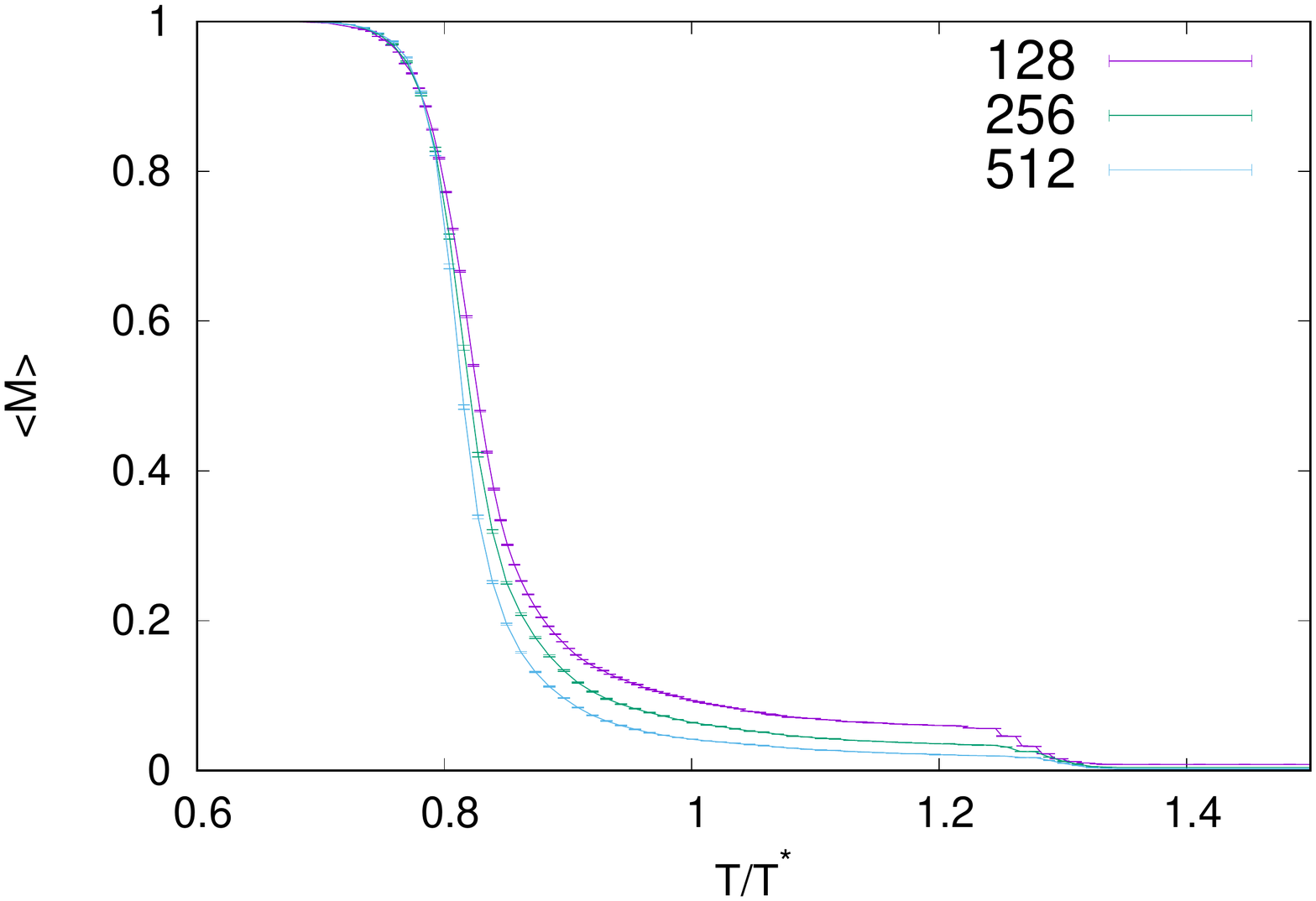}
  \end{center}
  \caption{Magnetisation for $\Delta=0.75$ and at the border $\sigma=1$}
  \label{fig:M_0.75_1}
\end{figure}

(plus signs on Fig.\ref{fig:phasediag}) At the point $\sigma=0.4$ in Fig.\ref{fig:M_0.75_0.4}
the average magnetization has a finite jump of $\Delta m \approx 1$ for all finite systems. The finite-size transition points, which
are identified with the position of the jump, $T_c(L)$, are shifted such that $T_c(L_1)<T_c(L_2)$ for $L_1<L_2$. Furthermore the distance
from the true transition point is well described by the asymptotic behaviour in the non-random system:
$\Delta T_c=T_c-T_c(L) \sim L^{-\sigma}$, since $T_C(L) \propto {\rm cst ~} \zeta_L(1+\sigma)$.
Thus the scaling exponent associated to lengths is $\nu \approx 1/\sigma$. At this point,
and in more general in the regime $\sigma \lesssim 0.5$ there is a \textit{random first-order transition} due to LR forces.

At the borderline value of $\sigma=0.5$ the magnetization profiles in Fig.\ref{fig:M_0.75_0.5} show still a jump,
at least for smaller finite systems. With increasing $L$, however, the jump in the magnetization is going to be rounded, so that the
transition could be continuous in the thermodynamic limit. With the finite-size results at hand we can not discriminate between these
scenarios. The shift of the finite-size transition points are characterized by an exponent:
$\nu \approx 2 = 1/\sigma$, in this case, too.

\paragraph{\underline{$0.5 < \sigma \le 1.0$: Mixed-order transitions}}
\label{sec:mixed}
(crosses in Fig.\ref{fig:phasediag}) In this regime we have a series of points with $\sigma=0.625,~0.75,~0.875$ and $1.0$ and the corresponding
profiles are shown in Figs.\ref{fig:M_0.75_0.625}-\ref{fig:M_0.75_1}.
The new feature of the profiles, that for different sizes they cross each other, so that for $T<T_c$ ($T>T_c$) the
profiles satisfy $m(L_1,T)<m(L_2,T)$ ($m(L_1,T)>m(L_2,T)$) for $L_1<L_2$. Furthermore at the transition point in the thermodynamic limit
the magnetization has a finite limiting value: $\lim_{L \to \infty} m(L,T_c^-)=m^->0$, which is different from the limit
$\lim_{L \to \infty} m(L,T_c^+)=m^+$. Consequently at the transition point there is a jump in the magnetization:
$\Delta m=m^--m^+$. We also expect that the actual value of $m^+$ is (close to) zero for strong disorder (large $\Delta$) and
it is increasing for smaller value of $\Delta$. In the thermodynamic limit for $T<T_c$ the magnetization
is expected to follow a singular temperature dependence:
$m(T)-m^- \sim (T_c-T)^{\beta}$. This can be checked in finite systems by defining finite-size transition points as the crossing
points of the profiles $m(L_1,T)$ and $m(L_2,T)$: $m[L_1,T_c(L_1,L_2)]=m[L_2,T_c(L_1,L_2)] \equiv m^-(L_1,L_2)$. According to scaling
theory the differences should behave asymptotically as: $T_c(L_1,L_2)-T_c \sim (L_1 L_2)^{-\frac{1}{2 \nu}}$ and
$m^--m^-(L_1,L_2) \sim  (L_1 L_2)^{-\frac{\beta}{2 \nu}}$. Due to strong finite-size corrections we could make an estimate for the critical
exponents only in the case $\sigma=0.875$ with the result: $1/\nu \approx 1.27$ and $\beta/\nu \approx 0.78$. This means that in this point,
or more generally in the $0.5 < \sigma \le 1.0$ part of the phase diagram (with $\Delta=0.75$) there is a mixed-order phase-transition
in the system: the magnetization has a jump at the transition point, but the correlation length is divergent at $T_c$.

Comparing the magnetization profiles at different values of $\sigma$, one can notice, that its limiting value, $m^-$, and thus the jump
$\Delta m$ is an increasing function of $\sigma$ in the given range. Increasing $\sigma$ over the upper limit, $\sigma=1$,
the form of the singularity changes ones more.

\paragraph{\underline{$\sigma > 1.0$: SR first-order transitions}}
\label{sec:SR first}

\begin{figure}
  \begin{center}
    \includegraphics[width=9cm]{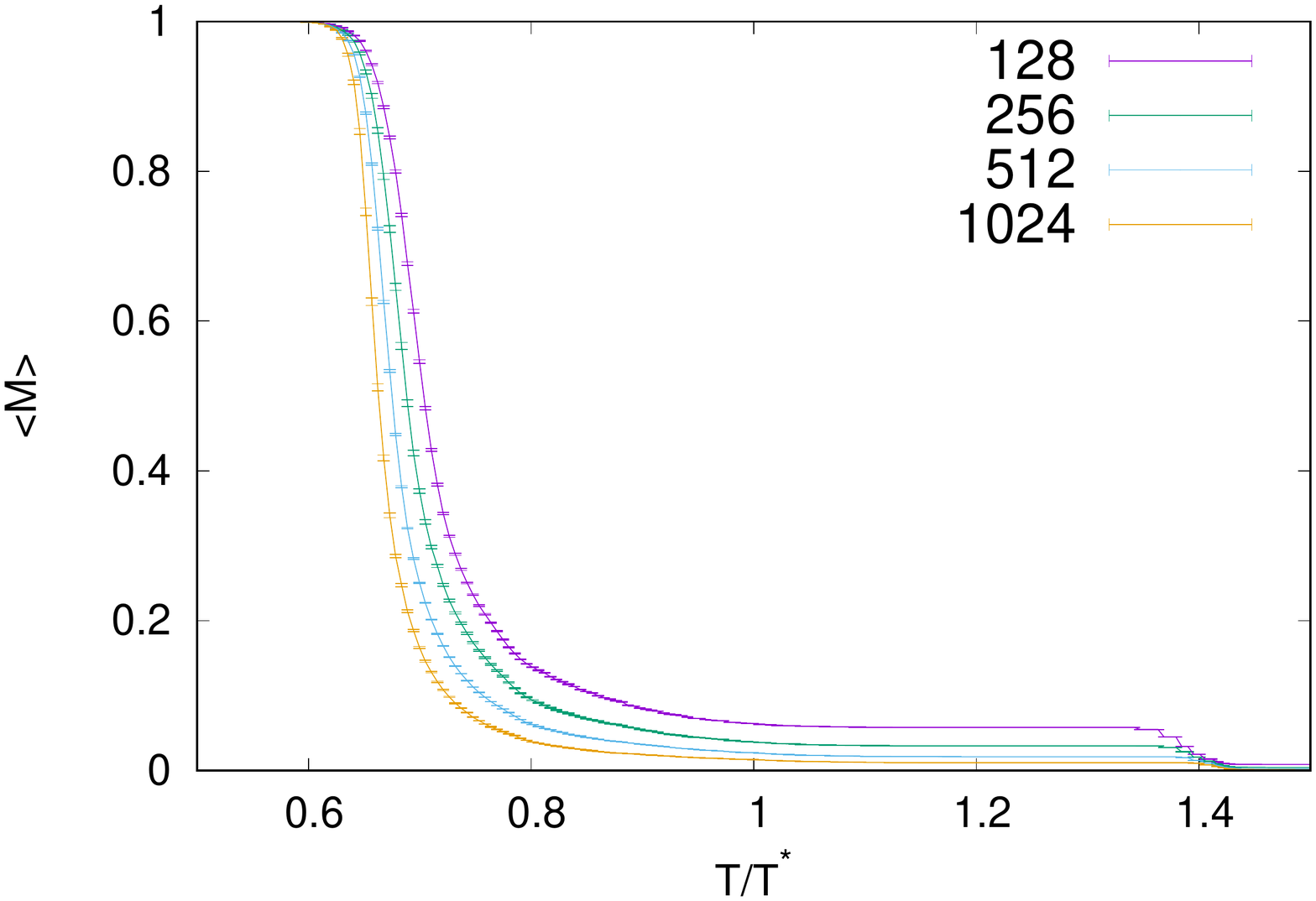}
  \end{center}
  \caption{First-order transition due to SR forces at $\Delta=0.75$ and $\sigma=1.25$}
  \label{fig:M_0.75_1.25}
\end{figure}

\begin{figure}
  \begin{center}
    \includegraphics[width=9cm]{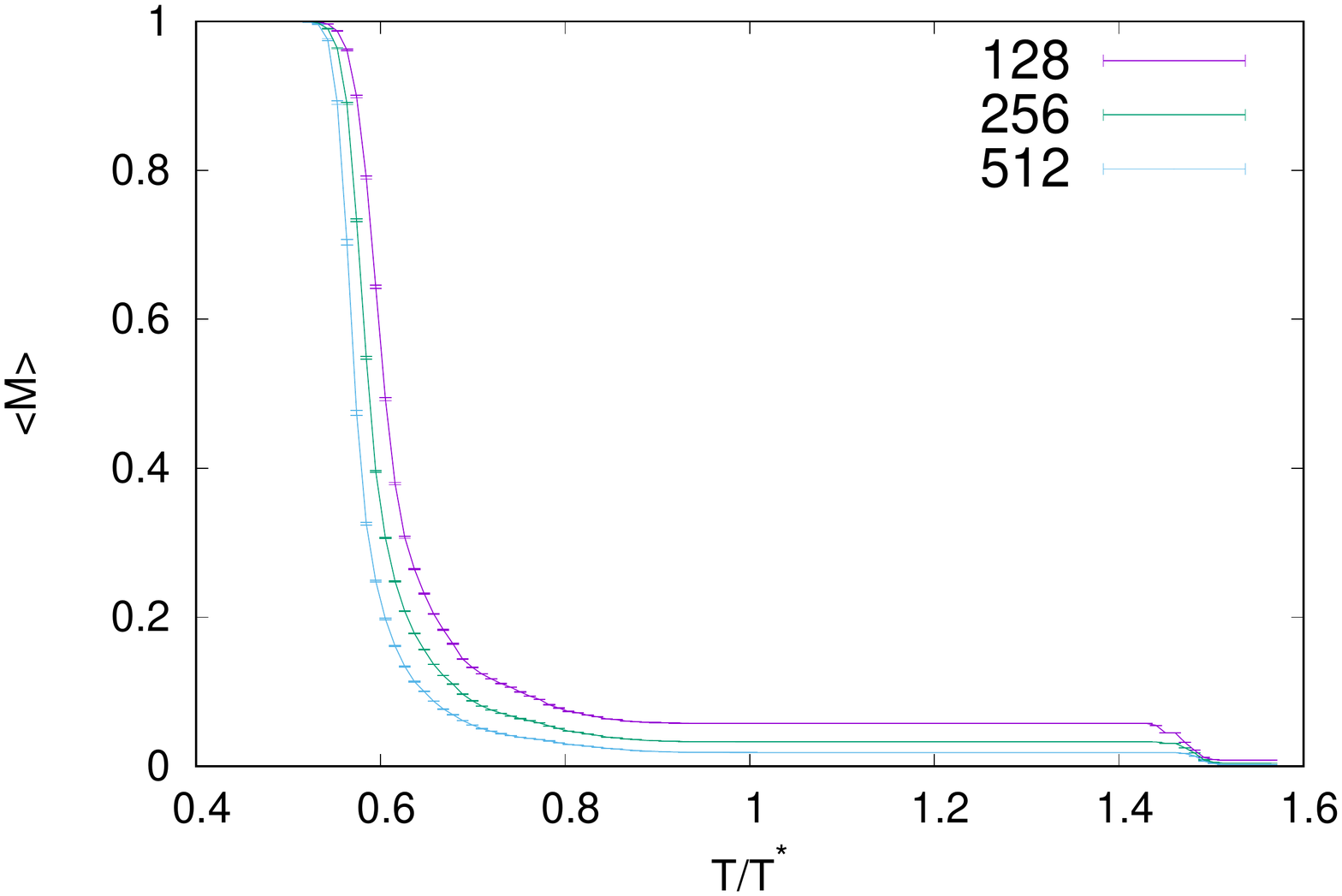}
  \end{center}
  \caption{First-order transition due to SR forces at $\Delta=0.75$ and $\sigma=1.5$}
  \label{fig:M_0.75_1.5}
\end{figure}

(circles on Fig.\ref{fig:phasediag}) The magnetization profiles at $\sigma=1.25$ and $1.5$ in Fig.\ref{fig:M_0.75_1.25} \ref{fig:M_0.75_1.5} show similar features: a jump is developed for large $L$, the asymptotic
position of which is at $T_c(\sigma)/\zeta(1+\sigma)<1$, which ratio is decreasing with increasing $\sigma$ and in the true SR model
with $\sigma \to \infty$ this ratio is just $1-\Delta$. Thus in this region the transition is of first order due to SR interactions. Comparing
the finite-size transition temperatures, $T_c(L)$, defined as the inflection point of the profiles, we observe the asymptotic behaviour:
$T_c(L)-T_c \sim L^{-1}$, characteristic for SR forces. We note that spontaneous order in the LR Potts chain for $\sigma > 1$ can be observed only in the $q \to \infty$ limit. For any finite value of $q$ due to thermal fluctuations there is no ordered phase, thus the
SR first-order transition regime is absent.


\section{Discussion}
\label{sec:discussion}
We have studied numerically the phase-diagram of the ferromagnetic LR Potts chain with random nearest-neighbour couplings in the $q \to \infty$ limit.
Depending on the strength of disorder, $\Delta$, and the decay exponent, $\sigma$, different type of phase-transitions are found:
first-order due to LR interactions, first-order due to SR interactions, second-order and mixed-order transitions. A schematic phase-diagram
is depicted in Fig.\ref{fig:phasediag}.


For small values of $\sigma < \sigma_c(\Delta) \le 0.5$ the long-range interactions are dominant over quenched disorder and the transition is of first order, as in the non-random system. For large values of $\sigma > 1.$ the transition is also of first-order, however now due to
short-range interactions. We note, that for finite-values of $q$ in this region there is no ferromagnetic order in the system.
For intermediate values of the decay exponent: $\sigma_c(\Delta) < \sigma < 1.$ quenched disorder is going to change the order of the transition.
For weaker disorder the transition turns to second-order, which is manifested by a divergent specific heat and by a divergent correlation length,
however the magnetization at the critical point is continuous and has a finite value. For strong disorder the transition turns to be of
mixed-order. At the transition point the correlation length is divergent, but there is a finite jump in the magnetization, as well as in
the energy-density. The finite-size scaling behaviour of the magnetization profiles are also different in the SO and the MO transitions.

The different type of transitions are connected with the geometric properties of the optimal graphs. At first-order transitions the optimal
graphs are different at the two sides of the transition points: in the ferromagnetic phase there is a giant cluster, whereas in the
high-temperature phase the clusters have finite mass and extent. At the second-order transition at both sides there is a giant cluster,
however at the transition a hole in this giant cluster is developed, the size of which as well as its mass is divergent. This hole
in the SO transition point is a fractal, therefore the average magnetization is continuous.
Similar process takes place at a mixed-order transition, too, with the difference, that in this case the ``hole'' in the high-temperature phase 
is a compact object having a finite density of mass. This leads to a jump in the magnetization in the thermodynamic limit. For large enough
$\Delta$ this hole is
going to disconnect the giant cluster, so that the density of its mass, being the magnetization has a vanishing value in the thermodynamic limit.

We expect that the results summarized in the phase-diagram in Fig.\ref{fig:phasediag} remain qualitatively correct for another, more general models, too. First we mention, that the LR forces in Eq.(\ref{hamiltonian}) can be (weakly) random, too, which means that in Eq.(\ref{J(r)}) the prefactor is modified as $J \to J_i$, and the $J_i > 0$ are random variables. Another set of models are obtained if the
parameter $q$ is a large, but finite value. As noted before this model for $\sigma > 1$ has no ordered phase, however similar phase-diagram is expected to hold in the regime $0 < \sigma < 1$. This conjecture is based on the known results in the SR models, in which the properties of the phase transitions in different dimensions are found to be a smooth function of $q$, so that the $q \to \infty$ limit is not singular\cite{jacobsenpicco,ai03,long2d}. Further numerical work is needed to clarify, if similar relation holds also for the LR model.

\section*{Appendix}
\label{appendix}
In this appendix we give the solution of the optimal cooperation
problem on the two lines $\Delta=0$ and $\sigma \rightarrow \infty$
of the phase diagram. Let's recall that an optimal set is a
set of edges which maximizes the objective function
\begin{equation*}
  f(S;\beta) = c(S) + \beta \sum_{e \in S} J(e)
\end{equation*}
where $c(S)$ is the number of connected components of $S$ and
$\beta$ is the inverse temperature.

For any sample, the optimal set for zero
temperature is the set of all the bonds, while for high temperature
the optimal set is empty. Between these two limits the optimal set
changes at a finite number $n_T$ of temperatures ($n_T<L$). We call
these temperature breaking temperatures. If there is only one breaking
temperature ($n_T=1$) the model is maximally first order since the
magnetization jumps from zero to one.

Let us first consider the case $\Delta = 0$, therefore a non disordered model.
We show below that for any decreasing weight function of the distance
(as for example $d^{-(1+\sigma)}$) there is a single breaking temperature for any
size $L$. Note first that if a bond of length $d$ belongs to an optimal
set, then there is an optimal set to which {\it all} bonds of length $d$
are present. Indeed the permutation of the sites $i \rightarrow i+1$
preserve the length of the bonds and therefore any bonds of length $d$
belongs to some optimal set. Since the union of two optimal sets is also
an optimal set, we deduce that there is an optimal set to which all
bonds of length $d$ belong.
Suppose now that the bond between site 0 and site $d$ belongs to the optimal
set. Then the bond between the sites $d$ and $2d$ also belongs to the optimal
set and consequently the site $0$, $d$, $2d$ belongs to the same cluster.
More generally all the bonds between $\alpha d$ and $(\alpha+1)d$
also belong to the optimal set and consequently all sites $\alpha d$, where
the product is modulo $L$ and $\alpha$ an arbitrary integer,
belong to the same cluster . If $L$ is a prime number, then all the sites
will be attained, and therefore the optimal set contain all bonds if it
contains any one, which proves the results. Note that in this special case
of $L$ being prime we did not use the fact that the weight function
is decreasing.
To sketch the results in the case where $L=ln$ is not a prime number
we introduce the sets of edges $C_{n,l}(k)$ induced by the
vertex sets $\left\{k,l+k,2l+k,\cdots,\left(n-1\right)l+k\right\}$.
It is clear that every optimal set is of the form
$R(n)=\bigcup_{k=0}^{l-1}C_{n,l}(k)$ and is therefore characterized by a
divisor of $L$. Showing that the
transition is maximally first order amounts to showing that the optimal
set is characterized by only either $1$ or $L$.
To this end, let's introduce the sets of edges $\Gamma_{n}(k)$ induced
by the set of vertices
$\left\{k,k+1,\cdots,k+\left(n-1\right)\right\}$. A union
$S(n)=\bigcup_{k=0}^{l-1}\Gamma_{n}(nk)$ is in general not a optimal set.
However comparing the objective function for $S(n)$ and $R(n)$
and using the fact that $J$ is a decreasing function of the distance,
we find that only $S(1)$ and $S(L)$ can be the optimal set. This proves
that the model is maximally first order also when $L$ is not prime.

\medskip
Now we turn to the case $\sigma \rightarrow\infty$,
{\it ie} when only the short range disordered bonds are present.
In the general case the couplings constant can take $n$ values
$0 < J_0 \le J_1 \le \cdots \le J_{n-1}$ the breaking temperatures
$T_k = \frac{1}{k-1}\sum_{i=0}^{L-1}J_i$ for $2 \le k \le L$.
Using this relation in the case of bimodal distribution with an
equal number of strong ($1+\Delta$) and weak ($1-\Delta$) bonds,
one gets
$1-\Delta = J_0 = \cdots =J_{\frac{L}{2}} < J_{\frac{L}{2}}=\cdots=J_{L-1}=1+\Delta$
from with the $T_k$ are easily deduced. After some algebra one gets that
if $\Delta \le \frac{1}{L-1}$ the model is maximally first order with
a breaking temperature $\frac{L}{L-1}$, while if $\frac{1}{L-1}<\Delta$
there are two breaking temperatures $T_1 = \frac{L}{L-2}(1-\Delta)$ and
$T_2 = 1+\Delta$. In the intermediate regime $T_1 \le T \le T_2$
the free energy is $f(T,L)=\frac{1}{2} +\frac{1}{2} \frac{1+\Delta}{T}$,
and we have numerically observed that  magnetization scales as $L^{-0.82}$.
So in the thermodynamical limit the magetization jumps from 0 to 1 at $T_2$.

Note finally that {\it all} realizations have exactly the same behavior.
Therefore in some sense, the model is not disordered.

\begin{acknowledgments}
This work was supported by the Hungarian Scientific Research Fund under grant
No. K109577 and K115959. J-Ch Ad’A extends thanks to the "Theoretical Physics Workshop" and FI to the Universit\'e Joseph Fourier for
supporting their visits to Budapest and Grenoble, respectively. 
\end{acknowledgments}

\end{document}